\documentstyle[12pt]{article}
\def\caja{\mathsurround=0pt}
\def\eqalign#1{\,\vcenter{\openup1\jot \caja
        \ialign{\strut \hfil$\displaystyle{##}$&$
        \displaystyle{{}##}$\hfil\crcr#1\crcr}}\,}
\def\etal{\hbox{\it et. al.}}
\def\bra#1{\left\langle #1\right|}
\def\ket#1{\left| #1\right\rangle}

\def\holdtheequation{\arabic}
\def\sectioneq{\def\holdtheequation{\thesection.\arabic}\let
      \section={\section\setcounter{equation}{0}}\setcounter
      {equation}{0}\def\theequation{\holdtheequation{equation}}}

\def\auto{\eqno(\refstepcounter{equation}\theequation)}

\def\supercite{\def\cite{\newcite}}

\def\newcite#1{\super{\newcount\citenumber\citenumber=0\getcite#1,@, }}
\def\getcite#1,{\advance\citenumber by1
\def\getcitearg{#1}\def\lastarg{@}
    \ifnum\citenumber=1
    \ref{#1}\let\next=\getcite\else\ifx\getcitearg\lastarg\let\next=\relax
    \else ,\ref{#1}\let\next=\getcite\fi\fi\next}
\def\zp{Z.\ Phys.\ }
\def\np{Nucl.\ Phys.\ }
\def\pr{Phys.\ Rev.\ }
\def\prl{Phys.\ Rev.\ Lett.\ }
\def\pl{Phys.\ Lett.\ }
\renewcommand{\thefootnote}{\fnsymbol{footnote}}
\def\mainhead#1{\setcounter{equation}{0}\addtocounter{section}{1}
     \vbox{\begin{center}\large\bf #1\end{center}}\nobreak\par}
\def\tablehead#1{\vbox{\begin{center}\large\bf #1\end{center}}\nobreak\par}
\def\subhead#1{\vbox{\medskip\noindent \bf #1}\nobreak\par}
\def\srf#1{$^{#1}$\ }
\def\rf#1#2#3{{\bf #1}, #2 (19#3)}

\supercite \sectioneq

\begin{document} \begin{titlepage} \rightline{\vbox{\halign{#\hfil \cr
\normalsize ANL-HEP-PR-94-90 \cr
\normalsize December 5, 1995 \cr}}}
\vspace{.75in}
\begin{center}

\Large
{\bf What We Can Learn About Nucleon Spin Structure From Recent Data}
\medskip\medskip\medskip

\normalsize M. Goshtasbpour
\\ \smallskip
{\it Center for Theoretical Physics and Mathematics, AEOI, \\
P.O. Box 11365-8486, Tehran, Iran}
\\ \smallskip
and
\\ \smallskip
{\it Department of Physics, Shahid Beheshti University, Tehran, Iran}
\\ \smallskip
and
\\ \smallskip
Gordon P. Ramsey
\footnote{Work supported in part by the U.S. Department of Energy, Division of
High Energy Physics, Contract W-31-109-ENG-38.}
\\ \smallskip
{\it Physics Department, Loyola University Chicago, Chicago, IL 60626}
\\ \smallskip
and
\\ \smallskip
{\it High Energy Physics Division, Argonne National Laboratory,
Argonne, IL 60439} \end{center}

\begin{abstract}
We have used recent data from CERN and SLAC to extract information about
nucleon spin structure. We find that the SMC proton data on $\int_0^1
\>g_1^p \>dx$, the E142 neutron data on $\int_0^1 \>g_1^n \>dx$ and the
deuteron data from SMC and E143 give different results for fractions of the
spin carried by each of the constituents. These appear to lead to two different
and incompatible models for the polarized strange sea. The polarized gluon
distribution occuring in the gluon anomaly does not have to be large in order
to be consistent with either set of experimental data. However, it appears that
the discrepancies in the implications of these data cannot be resolved with any
simple theoretical arguments.
We conclude that more experiments must be performed in order to adequately
determine the fraction of spin carried by each of the nucleon constituents.
\end{abstract}

\renewcommand{\thefootnote}{\arabic{footnote}} \end{titlepage}

\mainhead{I. Introduction}

One of the important questions in high energy physics is how nucleon spin
is related to the spins of the quark and gluon constituents.
Significant interest in high energy polarization was piqued a few years ago
when the European Muon Collaboration (EMC)\srf{1} analyzed polarized
deep-inelastic scattering (DIS) data which appeared to contradict theoretical
predictions, creating the "spin crisis". Since then, a flurry of theoretical
and experimental work has been performed to address this "crisis" and
further investigate the spin properties of the nucleons.

The spin dependent asymmetry in the deep-inelastic scattering of longitudinally
polarized leptons on longitudinally polarized nucleons is given by
$$
\eqalign{
A&=\Biggl[{{\sigma(\leftarrow \rightarrow)-\sigma(\leftarrow \leftarrow)}
\over {\sigma(\leftarrow \rightarrow)+\sigma(\leftarrow \leftarrow)}}\Biggr]
\approx D\cdot A_1,}
\auto\label{1.1}
$$
where the arrows refer to the relative longitudinal spin directions of the
beam and target, respectively, and it is assumed that $A_2$ is small, since it
is bounded by $R=\surd {{\sigma_L}\over {\sigma_T}}$. Information about the
polarized quark distributions can be extracted from this asymmetry by
$$
\eqalign{
A_1={{\sum_{\scriptstyle i} e_i^2\Delta q_i(x)}\over {\sum_{\scriptstyle i}
e_i^2 q_i(x)}},} \auto\label{1.2}
$$
where the sums are over all quark flavors. The proton structure function
$g_1^p$ can be extracted from the asymmetry $A_1$ by using
$$
\eqalign{
g_1^p(x,Q^2)\approx {{A_1(x)\>F_2(x,Q^2)}\over {2x(1+R)}}.} \auto\label{1.3}
$$
It is assumed that the transverse structure function $g_2^p$ is small and that
$A_1$ is relatively independent of $Q^2$, which has been verified by
experimental measurements. The extrapolation of the EMC data for $g_1^p$ to
lower Bjorken $x$ led to implications that, although the Bjorken sum rule (BSR)
of QCD\srf{2} was satisfied, the Ellis-Jaffe sum rule\srf{3}, based on a simple
quark model, was violated. Recently, the Spin Muon Collaboration (SMC) group
from CERN\srf{4} and the E142/E143 experimental groups from SLAC\srf{5} have
measured $A_1^p$ and $g_1^p$ to even lower $x$ values and have added the
corresponding neutron and deuteron structure functions $A_1^n$, $g_1^n$,
$A_1^d$, and $g_1^d$. These groups have also improved statistics and lowered
the systematic errors from the original data.

The DIS experiments with equations (1.2) and (1.3) can provide a means by
which we can extract the polarized quark distribution functions. We can
check the consistency of these distributions by comparing proton data
($g_1^p$), neutron data ($g_1^n$) and deuteron data ($g_1^d$) via sum rules.
There are many possibilities for models of the polarized quark and gluon
distributions which are consistent with sum rule and data constraints.
The motive here is to point out some of these possibilities and compare our
analysis to that of others. For example, Close and Roberts\srf{6} have done an
analysis of the proton and neutron DIS data with an emphasis on the integrated
distributions and the overall flavor contributions to nucleon spin. Ellis and
Karliner\srf{7} have done a similar analysis which includes higher order QCD
corrections. We have done a more detailed flavor dependent analysis including
the QCD corrections and the effect of the gluon anomaly. We will proceed by
assuming that the polarized gluon distribution is of moderate size and find
that the resulting polarized quark distributions which are consistent with
data and the appropriate sum rules. Our approach to this analysis consists of
three parts: (1) separating the valence and sea integrated parton
distributions for each flavor using different data sets as a basis to perform
the analyses, (2) discussing similarities and differences between the
phenomenological implications of the experimental results, and (3) suggesting
a set of experiments which would distinguish the quark and gluon contributions
to the proton spin. Our analysis differs from that of the experimental groups
in that we use sum rules in conjunction with a single experimental result to
extract the spin information, while they use data from multiple experiments
in order to check the validity of the sum rules. In addition, they assume a
flavor symmetric sea and ignore anomaly contributions. However, there is
relatively good physics agreement with our results and those of the
experimental groups.

The paper is structured as follows: In part II, we discuss the theoretical
basis for determining the polarized parton distributions and the assumptions
we have made to generate them. In part III, we discuss our phenomenological
analysis of the existing data and the consistency of the various models.
Part IV is a discussion of the experiments which can be performed with existing
accelerators to further our knowledge of the spin content of nucleons.

\mainhead{II. Theoretical Background}

\subhead{A. Polarized Quark Distributions}

Fundamentally, we assume that the nucleons are comprised of valence quarks,
whose polarized and integrated distributions are defined by:
$$
\eqalign{
\Delta q_v (x,Q^2)&\equiv q_v^+(x,Q^2)-q_v^-(x,Q^2) \cr
\langle \Delta q_v (Q^2)\rangle&\equiv \int_0^1 \Delta q_v(x,Q^2)\>dx,}
\auto\label{2.1}
$$
where $+(-)$ indicates the quark spin aligned (anti-aligned) with the nucleon
spin. In order to construct the polarized quark distributions from the
unpolarized ones, we can start with a modified 3-quark model based on an SU(6)
wave function for the proton. This model is based on flavor symmetry of the
u- and d-sea and constructs the valence distributions to satisfy the Bjorken
sum rule.\srf{8} The valence quark distributions can be written in the form:
$$
\eqalign{
\Delta u_v (x,Q^2)&=\cos \theta_D [u_v(x,Q^2)-{2\over 3}d_v(x,Q^2)], \cr
\Delta d_v (x,Q^2)&=-{1\over 3}\cos \theta_D d_v(x,Q^2),}\auto\label{2.2}
$$
where $\cos \theta_D$ is a "spin dilution" factor which vanishes as $x\to 0$
and becomes unity as $x\to 1$, characterizing the valence quark helicity
contribution to the proton.\srf{8,9} Normally, the spin dilution factor is
adjusted to satisfy the Bjorken sum rule and to agree with the deep-inelastic
data at large $x$.

To generate the valence quark distributions, we use the higher order set of
GRV\srf{10} unpolarized distributions, evolved to the $Q^2$ scales of each
experiment. These agree with the MRS\srf{11} distributions for $x\geq 0.05$.
The spin dilution factor in equation (2.2) was determined from the BSR, which
we have assumed valid. The consistency of the resulting polarized valence
distributions was checked by comparing with the value for the ratio of
proton and neutron magnetic moments:
$$\eqalign{
{{\mu_p}\over {\mu_n}}={{2\langle \Delta u_v\rangle-\langle \Delta d_v\rangle}
\over {2\langle \Delta d_v\rangle-\langle \Delta u_v\rangle}}\approx
-{3\over 2}.}\auto\label{2.3}
$$
With our values $\langle \Delta u_v\rangle=1.00\pm 0.01$ and $\langle \Delta
d_v\rangle=-.26\pm 0.01$, both the BSR and magnetic moment ratio are satisfied.
This also yields a spin contribution from the valence quarks equal to $0.74\pm
0.02$, consistent with other treatments of the spin content of quarks.
\srf{12,13} The quoted errors arise from data errors on $g_A/g_V$, and the
differences in choice of the unpolarized distributions used to generate the
polarized valence quark distributions. The original analysis by Qiu, {\it et.
al.}\srf{8} effectively reached the same conclusion.

The polarization of the sea occurs by gluons that are emitted by gluon
Bremsstrahlung and by quark-antiquark pair creation. The corresponding
integrated polarized sea distribution is defined as:
$$
\eqalign{
\langle\Delta S(Q^2)\rangle\equiv\langle[\Delta u(Q^2)+\Delta\bar{u}(Q^2)+
\Delta d(Q^2)+\Delta \bar{d}(Q^2)+\Delta s(Q^2)+\Delta \bar{s}(Q^2)]\rangle,}
\auto\label{2.4}
$$
where the polarized sea flavors are defined analogous to the valence quarks.
It is assumed that the lightest flavors dominate the sea polarization, since
the heavier quarks should be significantly harder to polarize. Thus, we
assume that the quark and antiquark flavors are symmetric, but break the SU(6)
symmetry of the sea by assuming that the heavier strange quarks will be less
polarized.\srf{8} Then, the sea distributions are related as follows:
$$
\eqalign{
\Delta \bar{u}(x,Q^2)&=\Delta u(x,Q^2)=\Delta \bar{d}(x,Q^2)=\Delta d(x,Q^2)\cr
&=[1+\epsilon]\Delta \bar{s}(x,Q^2)=[1+\epsilon]\Delta s(x,Q^2).}
\auto\label{2.5}
$$
The $\epsilon$ factor is a measure of the increased difficulty in polarizing
the strange sea quarks.

In terms of the proton wave function, we can write the integrated distributions
as
$$
\eqalign{
\langle\Delta q_i s^{\mu}\rangle=\bra{ps} \bar{q}\gamma^{\mu}\gamma_5 q_i
\ket{ps}/2m,}\auto\label{2.6}
$$
where $s^{\mu}(p)$ is the axial four-vector which characterizes a spin
$1\over2$ particle and $m$ is the mass of the particle. The integrated
polarized structure function, $I^{p(n)}\equiv\int_0^1 g_1^{p(n)}(x)\>dx$,
is related to the polarized quark distributions by
$$
\eqalign{
I^{p(n)}=&{1\over 2}(1-\alpha_s^{corr})\langle\bigl[{{4(1)}
\over 9}\Delta u_v+{{1(4)}\over 9}\Delta d_v+{{4(1)}\over 9}(\Delta u_s+\Delta
\bar{u})+{{1(4)}\over 9}(\Delta d_s+\Delta \bar{d}) \cr
&+{1\over 9}(\Delta s+\Delta\bar{s})\bigr]\rangle.}\auto\label{2.7}
$$
The QCD corrections, characterized by $\alpha_s^{corr}$, have been caluclated
to $O(\alpha_s^4)$\srf{14} and are
$$
\eqalign{
\alpha_s^{corr}\approx ({{\alpha_s}\over {\pi}})+3.5833({{\alpha_s}\over
{\pi}})^2+20.2153({{\alpha_s}\over {\pi}})^3+130({{\alpha_s}\over {\pi}})^4,}
\auto\label{2.8}
$$
where the last term is estimated. The higher twist corrections calculated by
Stein, $\etal$\srf{15} are small enough to neglect at the $Q^2$ values of
the data. The QCD corrections have a much more significant effect in extracting
information from the data and sum rules. In fact, although the last correction
term in eq. (2.8) is estimated, its effect on the extracted numbers is less
than the significant figures which we report.

Thus, the data on $g_1$ allows the determination of a linear combination of
$\epsilon$ and the overall size of the polarized sea. This is not enough,
however, to determine all of the sea parameters. Additional constraints are
provided by the axial-vector current operators, $A_{\mu}^k$, whose matrix
elements for the proton define the coefficients, $a^k$, as\srf{16}
$$
\eqalign{
\bra{ps} A_{\mu}^k \ket{ps}=s_{\mu}a^k,} \auto\label{2.9}
$$
where the $a^k$ are non-zero for k=0, 3 and 8. These current operators are
members of an SU(3)$_f$ octet, whose non-zero elements give relations between
the polarized distributions and the measurable coefficients $a^k$.
The Bjorken sum rule relates the polarized structure function $g_1(x)$,
measured in polarized deep-inelastic scattering, to the axial vector current
$A_{\mu}^3$. The coefficient $a^3$ is measured in neutron beta decay and this
sum rule is considered to be a fundamental test of QCD. In terms of the
polarized distributions and our assumptions about the flavor symmetry of the
$u$ and $d$ polarized sea, the Bjorken sum rule can be reduced to the form:
$$
\eqalign{
\int_0^1 [\Delta u_v(x,Q^2)-\Delta d_v(x,Q^2)]\>dx=a^3,}\auto\label{2.10}
$$
which enables us to determine the valence distributions, as previously
mentioned. Furthermore, since the BSR relates $g_1^p$ and $g_1^n$, the DIS data
on $g_1$ (for p, n and d) can be used to set constraints on the polarized
sea distributions.

The coefficient $a^8$ is determined by hyperon decay, reflecting the other
baryon axial charges in the symmetry. A traditional analysis of hyperon decays,
yields two empirical constants: $D$ and $F$,\srf{6} which are related to the
polarized quark distributions by $a^8$. This relation can be written as:
$$
\eqalign{
a^8=\langle\bigl[\Delta u_v+\Delta d_v+\Delta u_s+\Delta \bar{u}+\Delta d_s+
\Delta\bar{d}-2\Delta s-2\Delta \bar{s}\bigr]\rangle=3F-D\approx 0.58\pm 0.02.}
\auto\label{2.11}
$$
Lipkin has pointed out that one must proceed with caution in using hyperon spin
structures, however, without a suitable hyperon spin model.\srf{17}
Fortunately, our analysis is not highly sensitive to the value of $a^8$.

The factor $a^0$ is related to the total spin carried by the quarks in the
proton. Assuming that the $u$ and $d$ flavors are symmetric in the polarized
sea, we can relate the non-zero axial currents and the structure function
$g_1^p$ in the flavor-independent form:
$$
\eqalign{
a^0\approx 9(1-\alpha_s^{corr})^{-1}\int_0^1 g_1^p(x)\>dx-{1\over 4}a^8-
{3\over 4} a^3.}\auto\label{2.12}
$$
Thus, the equations for the axial current coefficients give constraints
to the polarized quark distributions, from which we can attempt to extract
specific information about individual contributions to the overall proton
spin. Shortly after the EMC experiment, there were a number of theoretical
calculations which isolated the contributions of each of the flavors of the
polarized sea to the proton spin.\srf{8,18} All of these arrived at the
conclusion that the sea is negatively polarized, which is reasonable when one
analyzes the spin dependent forces which cause polarization of the sea from
valence quarks and gluons. Updated values for the flavors of polarized
distributions can be determined from the recent SMC and SLAC data.\srf{4,5}

One can impose theoretical constraints on the polarized strange sea\srf{8}
by assuming that
$$\eqalign{
\mid \int_0^1 \Delta s(x)\>dx \mid \le {1\over 3}\int_0^1 x\overline{s}(x)\>dx
\approx 0.005.}\auto\label{2.13}
$$
This "Valence Dominated Model" (VDM) is based on a mechanism where sea quarks
obtain their polarization through a localized interaction with the valence
quarks. This model provides a more restrictive limit on the size of the
polarized strange sea than the positivity constraint discussed by Preparata,
Ratcliffe and Soffer.\srf{19} The VDM model can be compared with the integrated
distributions extracted from the data to check for its validity.

\subhead{B. Polarization of Gluons}

The gluons are polarized through Bremsstrahlung from the quarks. The integrated
polarized gluon distribution is written as
$$
\eqalign{
\langle\Delta G\rangle=\int_0^1 \Delta G(x,Q^2)\>dx=\int_0^1
[G^+(x,Q^2)-G^-(x,Q^2)]\>dx,}\auto\label{2.14}
$$
where the $+(-)$ indicates spin aligned (anti-aligned) with the nucleon, as
in the quark distributions. We cannot determine {\it a priori} the size of the
polarized gluon distribution in a proton at a given $Q^2$ value. The evolution
equations for the polarized distributions, indicate that the polarized gluon
distribution increases with $Q^2$ and that its evolution is directly related to
the behavior of the orbital angular momentum, since the polarized quark
distributions do not evolve in $Q^2$ in leading order.\srf{20} Thus, one
assumes a particular form for the polarized gluon distribution for a given
$Q^2$ and checks its consistency with experimental data which are sensitive to
$\Delta G(x,Q^2)$ at a particular $Q_0^2$. Initial analyses of the EMC
data\srf{18} led to speculation that the integrated gluon distribution may be
quite large, even at the relatively small value of $Q^2=10.7$ GeV$^2$.

The model of $\Delta G$ that is used has a direct effect on the measured value
of the quark distributions through the gluon axial anomaly.\srf{21} In QCD,
the U(1) axial current matrix element $A_{\mu}^0$ is not strictly conserved,
even with massless quarks. Hence, at two loop order, the triangle diagram
between two gluons generates a $Q^2$ dependent gluonic contribution to the
measured polarized quark distributions. This term has the general form:
$$
\eqalign{
\Gamma (Q^2)={{N_f\alpha_s(Q^2)}\over {2\pi}}\int_0^1 \Delta G(x,Q^2)\>dx,}
\auto\label{2.15}
$$
where $N_f$ is the number of quark flavors. Thus, for each flavor of quark
appearing in the distributions, the measured polarization distribution is
modified by a factor: $\langle\Delta q_i\rangle-\Gamma(Q^2)/N_f$.
In order for us to determine the quark contributions to the spin of the
nucleons, it is necessary for us to know the relative size of the polarized
gluon distribution. If we base our analysis solely on the naive quark model,
then $\sum \Delta q \to 1$ and $\Delta G$ may be quite large to be
consistent with EMC data. This is surprising, since there are
no high-spin excited states of nucleons which create such a large $\Delta G.$
If we consider the polarized distributions of Qiu {\it et. al.}, a reasonably
sized $\Delta G$ is possible if the sea has a suitably negative polarization.

We have considered two possible models for $\Delta G$:
$$\eqalign{ (1)& \qquad \Delta G(x)= x\> G(x), \cr
(2)& \qquad \Delta G=0.} \auto\label{2.16} $$
The first implies that the spin carried by gluon is the same as its momentum,
motivated by both simple PQCD constraints and the form of the splitting
functions for the polarized evolution equations.\srf{20,22} The second provides
an extreme value for determining limits on the values of the polarized sea
distribution.

Another natural constraint to the polarized distributions relates the
integrated parton distributions to the orbital angular momentum of the
constituents. Due to O(2) invariance, a proton with momentum and spin in the
z-direction will conserve $J_z$. This total spin sum rule can be written in
terms of the polarized distributions as:
$$
\eqalign{
J_z={1\over 2}={1\over 2}\langle\Delta q_v\rangle+{1\over 2}\langle
\Delta S\rangle+\langle\Delta G\rangle+L_z.}\auto\label{2.17}
$$
The right hand side represents the decomposition of the constituent spins along
with their relative angular momentum, $L_z$. Although this does not provide
a strict constraint on either $\Delta q_{tot}$ or $\Delta G$, it does give
an indication of the fraction of total spin due to the angular momentum
component as compared to the constituent contributions.

\subhead{C. Interpolation of Data at Small-x}

All sets of data are limited in the range of Bjorken $x$ and thus, the
integrals must be extrapolated to $x\to 0.$ Thus, the possibility of existance
of a Regge type singularity at $x\to 0$ is not accounted for in the analyses.
A significant singularity could raise the value of $g_1^p$ towards the naive
quark model value and could account for some of the discrepancy between the
original EMC data and the Ellis-Jaffe sum rule.\srf{3} In light of the recent
HERA data,\srf{23} there is the possibility that the increase in $F_2$ at small
$x$, even at the lower $Q^2$ values of the E142/E143 data, could indicate a
change in the extrapolated values of these integrals. These possibilities are
a topic for future study. For the purposes of this paper, we will assume that
this overall effect of $F_2$ on $g_1^p$ will not alter the integral by any more
than the present experimental errors. We use the unpolarized distributions of
MRS and GRV in section III B since they include the small-$x$ data from HERA.
The shape of the polarized gluon distribution at small-$x$ affects the anomaly
term, and thus the overall quark contributions to the integrals. Future
experiments can shed light on the size of this effect, a detail we will discuss
in section IV. We believe that the present data show that anomaly effects are
limited and the overall integrated polarized gluon distribution is not very
large at these energies. This point is discussed in the next section.

\mainhead{III. Phenomenology}

\subhead{A. Assumptions and Analysis using New Data}

We consider recent SMC\srf{4} and E142/143\srf{5} data to extract polarization
information about the sea. The SMC experiment, which measured
$\int_0^1 g_1^p\>dx$, consisted of deep inelastic scattering of polarized muons
off of polarized protons in the kinematic range $0.003\le x\le 0.7$ and 1
GeV$^2 \le Q^2\le$ 60 GeV$^2$. The data were then extrapolated to yield
the integrated value of the structure function. In the other SMC experiment,
the polarized proton target was replaced by a polarized deuteron target and
$\int_0^1 g_1^d\>dx$ was extracted from data in the kinematic range
$0.003\le x\le 0.7$ and 1 GeV$^2 \le Q^2\le$ 60 GeV$^2$. The E142
experiment extracted $\int_0^1 g_1^n\>dx$ from data in the kinematic range
$0.03\le x\le 0.6$ and 1 GeV$^2 \le Q^2\le$ 60 GeV$^2$ by scattering polarized
electrons off of a polarized $^3$He target. The E143 experiment
measured $\int_0^1 g_1^n\>dx$ in the kinematic range $0.03\le x\le 0.8$ and
1 GeV$^2 \le Q^2\le$ 10 GeV$^2$ by scattering polarized electrons off of a
solid polarized deuterated ammonia $^{15}$ND$_3$ target. The integrated
results with errors and average $Q^2$ values are summarized in Table I.

\tablehead{Table I: Experimental Parameters for the Integrated Structure
Functions}
$$\begin{array}{ccccc}
 Quantity    & SMC\>(I^p) & SMC\>(I^d) & E142\>(I^n) & E143\>(I^d) \cr
 I^{exp}      &.136       &.034        & -.022       &  .041     \cr
 Stat.\>err.  & \pm .011  & \pm .009   & \pm .007    & \pm .003  \cr
 Sys.\>err.   & \pm .011  & \pm .006   & \pm .006    & \pm .004  \cr
 Avg.\>Q^2\>(GeV^2) & 10.0 & 10.0      & 2.0         & 3.0       \cr
 \alpha_s(Q^2) & .27      & .27        & .385        & .35
\end{array}$$

We can write the integrals of the polarized structure functions, $\int_0^1
g_1^i\>dx$ in the terms of the coefficients $a^k$:\srf{6}
$$
\eqalign{
I^p&\equiv \int_0^1 g_1^p(x) dx=\Biggl[{{a^3}\over {12}}+{{a^8}\over {36}}+
{{a^0}\over 9}\Biggr]\Bigl(1-\alpha_s^{corr}\Bigr), \cr
I^n&\equiv \int_0^1 g_1^n(x) dx=\Biggl[-{{a^3}\over {12}}+{{a^8}\over {36}}+
{{a^0}\over 9}\Biggr]\Bigl(1-\alpha_s^{corr}\Bigr), \cr
I^d&\equiv \int_0^1 g_1^d(x) dx=\Biggl[{{a^8}\over {36}}+{{a^0}\over 9}\Biggr]
\Bigl(1-\alpha_s^{corr}\Bigr)(1-{3\over 2}\omega_D),} \auto\label{3.1}
$$
where $\omega_D$ is the probability that the deuteron will be in a D-state.
Using N-N potential calculations, the value of $\omega_D$ is about $0.058$.
\srf{4,24} The difference $(I^p-I^n)$ is the Bjorken sum rule, which is
fundamental to the tests of QCD. There seems to be agreement in the
experimental papers that the data from each substantiates the Bjorken sum rule,
to within the experimental errors. We have assumed that the BSR is valid,
and have used it as a starting point for extracting an effective $I^p$ value
from neutron and deuteron data. The comparison of the effective $I^p$ values
gives a measure of the consistency of the different experiments to the BSR.

Considerable discussion regarding results of these measurements focuses on
the Ellis-Jaffe sum rule (EJSR),\srf{3} which predicts the values of $g_1^p$
and $g_1^n$ using an unpolarized sea. This has the form:
$$
\eqalign{
I^p={1\over {18}}\bigl[9F-D\bigr]\Bigl(1-\alpha_s^{corr}\Bigr), \cr
I^n={1\over {18}}\bigl[6F-4D\bigr]\Bigl(1-\alpha_s^{corr}\Bigr),}
\auto\label{3.2}
$$
where F and D are the empirically determined $\beta$-decay constants,
constrained so that their sum: $F+D=g_A/g_V$ satisfies the Bjorken sum rule.
Using the approximate values,\srf{6} $F\approx 0.46\pm 0.01$ and $D\approx
0.80\pm 0.01$, this sum rule predicts that $I^p=0.161$ and $I^n=-0.019$.
These values of $F$ and $D$ also yield $$a^8\equiv 3F-D=0.58\pm 0.03.
\auto\label{3.3}$$ Clearly, the E142 ($I^n$) data are consistent with this sum
rule, while the other data are not. Higher order corrections to the EJSR have
been calculated,\srf{25} but amount to about a 10$\%$ correction to the values
of the integrals and are not enough to account for the discrepancy with the
SMC data. Higher twist corrections\srf{26} are only significant at the lower
$Q^2$ values of the E142 $I^n$ data, where there is agreement with the EJSR.
The point thus focuses on the discussion of the size of the polarized sea,
which differs in analyses of these data. We address this in detail later.

The experimental values of $I^{exp}$ for the proton, neutron and deuteron,
combined with the value of $\Delta q_v$ determined in section II, can be used
to determine the polarization of each of the sea flavors. With anomaly
correction, the total spin carried by each of the flavors can be written as:
$$\eqalign{
\langle \Delta q_{i,{\rm val}}+\Delta q_{i,{\rm  sea}}+\Delta\overline{q}_i
-\frac{\alpha_s}{2 \pi} \Delta G\rangle =\langle \Delta q_{i,{\rm tot}}
\rangle.} \auto\label{3.4}$$
The anomaly terms included in the quark distributions have the form of equation
(2.15) using both models of the polarized glue from (2.16).

Since the anomalous dimensions for the polarized distributions have an
additional factor of $x$ compared to the unpolarized case, early treatments of
the spin distributions assumed a form of: $\Delta q(x)\equiv xq(x)$ for all
flavors.\srf{22} We have compared this form of the distributions to those
extracted from the recent data, using the defined ratio $\eta \equiv{{\langle
\Delta q_{sea}\rangle_{exp}}\over {\langle xq_{sea}\rangle_{calc}}}$ for each
flavor.

\subhead{B. Results for the Polarized Distributions}

The analysis for each polarized gluon model proceeds as follows:

(i) We extract a value of $I^p$ from either the data directly or via the
BSR using equation (3.1). Then, equation (2.12) is used to extract $a_0$.
The anomaly dependence on both sides of eq. (2.12) cancels, but the overall
contribution to the quark spin, $\langle \Delta q_{tot}\rangle=a_0+\Gamma$,
includes the anomaly term. The value $a_8$ from the hyperon data with equations
(2.11) and (2.12) are then used to extract $\Delta s$ for the strange sea.
The total contribution from the sea then comes from $\langle\Delta q_{tot}
\rangle=\langle\Delta q_v\rangle+\langle\Delta S\rangle$. The factor $\epsilon$
and the distributions $\langle \Delta u\rangle_{sea}=\langle \Delta d\rangle_
{sea}$ are then derived from equations (2.4) and (2.5). Finally, the $J_z$
= 1/2 sum rule (equation 2.17) gives $L_z$.

(ii) Since the VDM model is based upon the chiral distributions, we calculate
the corresponding results in Table IIa, where the anomaly term is zero. Here,
$\Delta s$ comes from the VDM assumption for the strange sea (equation 2.13).
Then $\langle\Delta q\rangle_{tot}=a_0$ can be extracted from
$a_0-a_8=6\langle\Delta s\rangle$. Finally, $g_1^p$ comes from (2.12), and the
other sea information can be extracted. This provides a theoretical limit on
these quantities, based upon a restricted strange sea polarization.

The overall results are presented in Table II ($\Delta G=xG$) and Table IIa
($\Delta G=0$). The E143 proton data\srf{5} gives virtually the same numbers
as the deuteron data shown in these tables.

\tablehead{Table II: Integrated Polarized Distributions: $\Delta G=xG$}
$$\begin{array}{cccccc}
  Quantity    &SMC(I^p)  &SMC(I^d)  &E142(I^n)  &E143(I^d) \cr
                                                       \cr
  <\Delta u>_{sea} &-.077  &-.089   &-.050      &-.068 \cr
  <\Delta s>       &-.037  &-.048   &-.010      &-.028 \cr
  <\Delta u>_{tot} &0.85   &0.82    &0.90       &0.87  \cr
  <\Delta d>_{tot} &-.42   &-.43    &-.36       &-.40  \cr
  <\Delta s>_{tot} &-.07   &-.10    &-.02       &-.06  \cr
  \eta_u = \eta_d  &-2.4   &-2.8    &-1.6       &-2.1  \cr
    \eta_s      &-2.0    &-3.0      &-0.7       &-1.6  \cr
   \epsilon     &1.09    &0.84      &4.00       &1.41  \cr
    \Gamma      &0.06    &0.06      &0.08       &0.08  \cr
     I^p        &0.136   &0.129     &0.137      &0.131 \cr
  <\Delta q>_{tot} &0.36   &0.29    &0.52       &0.41  \cr
  <\Delta G>    &0.46    &0.46      &0.44       &0.44  \cr
     L_z        &-.14    &-.11      &-.20       &-.15
\end{array}$$

\tablehead{Table IIa: Integrated Polarized Distributions: $\Delta G=0$}
$$\begin{array}{cccccc}

 Quantity & SMC(I^P) & SMC(I^d) & E142(I^n) & E143(I^d) & VDM \\
   \\
 <\Delta u>_{sea} & -.087   & -.099  & -.063  &-.082  & -.045 \cr
 <\Delta s>       & -.047   & -.058  & -.023  &-.042  & -.005 \cr
 <\Delta u_{tot}> & .83     & .80    & .88    & .84   & .91   \cr
 <\Delta d_{tot}> & -.44    & -.45   & -.39   & -.43  & -.35  \cr
 <\Delta s_{tot}> & -.09    & -.12   & -.05   & -.08  & -.01  \cr
   \eta_u=\eta_d  & -2.7    & -3.2   & -2.0   & -2.6  & -1.4  \cr
    \eta_s        & -2.4    & -3.7   & -1.7   & -2.5  & -0.3  \cr
    \epsilon      & 0.86    & 0.70   & 1.71   & 0.96  & 8.00  \cr
    \Gamma        & 0.00    & 0.00   & 0.00   & 0.00  & 0.00  \cr
     I^p          & .136    & .129   & .137   & .131  & .152  \cr
 <\Delta q>_{tot} & 0.30    & 0.23   & 0.44   & 0.33  & 0.55 \cr
     L_z          & 0.35    & 0.39   & 0.28   & 0.35  & 0.23
\end{array}$$

{}From tables II and IIa, it is obvious that the naive quark model is not
sufficient to explain the characteristics of nucleon spin. However, these
results have narrowed the range of constituent contributions to the proton
spin. The following conclusions can be drawn, which lead to a modified view
of the proton's spin picture.

(1) The total quark contribution to the proton spin is between $1\over 4$ and
$1\over 2$, as opposed the quark model value of one, or the extracted EMC value
of zero. The errors in determining the total quark contribution are due to
experimental errors ($\pm 0.04\to \pm 0.08$ for $\Delta q_{tot}$) and the
uncertainty in the value for $\Gamma$ ($\approx \pm 0.04$). For a given
gluon model, however, the differences are slightly larger than one standard
deviation. This cannot be accounted for by different $Q^2$ values since the
polarized quark distribution does not evolve with $Q^2$ to leading order.
Also, higher twist effects do not appear to be large enough to account for this
difference.\srf{15} However, all of these values imply that the total sea,
and hence the strange sea, has a smaller polarization than originally thought
after the EMC experiment. The results for the strange sea in the proton and
deuteron data are still larger than the positivity bound of Preparata,
Ratcliffe and Soffer.\srf{19} There is an implication here that their
positivity bound value is too small, since it is based entirely on unpolarized
data. Nevertheless, all data imply that the strange quark spin contribution is
much smaller than that of the lighter quarks. The flavor symmetry is broken by
the large values of $\epsilon$, namely $0.7\le \epsilon \le 4.0$, as
opposed to $\epsilon=0$. It is also interesting to note that the results
obtained from the SMC proton data are consistent with a recent lattice QCD
calculation of the polarized quark parameters.\srf{12} Although the flavor
contributions to the proton spin cannot be extracted exactly, the range of
possibilities has been substantially decreased by these experiments.
Specifically, the up and down contributions agree to within a few percent.
The main difference remains the question of the strange sea spin content.

(2) Despite the differences, there are similarities among these sets of data.
All of the extracted values for $I^p$ are well within the experimental
uncertainties, indicating a strong agreement about the validity of the Bjorken
Sum Rule. We have arrived at this conclusion by using the BSR to extract $I^p$,
as opposed to the experimental groups, which used data to extract the BSR.
There seems to be general agreement as to the consistency of these results.

(3) As we mentioned in section II, the validity of the Ellis-Jaffe sum rule
reduces to the question of the size of the polarized sea. It also addresses
the assumptions made by the naive quark model and early polarization
calculations based on a simple SU(6) model for the proton. The sea results, the
deviation of $\epsilon$ from zero and the differences in $\eta$ from one, all
indicate that the models for quark polarizations must be modified to account
for the experimental results. The physical conclusions are: (i) that the
sea is polarized opposite to that of the valence quarks (see ref. 8 for
interpretations), (ii) that the strange quarks must be treated separately in
determining their contribution to the proton spin due to mass effects, and
(iii) that polarized distributions for each quark flavor must be modified
so that $\Delta q_f\approx \eta_f\>x\>q_f$, where $\eta$ is extracted from data
and is likely different from unity. Thus, the relation between unpolarized and
polarized distributions is likely more complex than originally thought.

(4) This analysis implies that the role of the anomaly correction
is significant only in the sense that minimizing errors in specifying spin
contributions from quark flavors depends on determining the size of the
polarized glue. By comparing the results given in tables II and IIa,
where analysis of the data is done with both a zero and a small anomaly
correction, we see the key results and conclusions are not significantly
different. Further, even if there are higher twist corrections to the anomaly
at small $Q^2$, this will not reconcile differences in the flavor dependence of
the polarized sea. The anomaly term does not vary significantly enough for the
$Q^2$ range of the data to explain any differences. However, the analysis of
$I^n$ with the anomaly in equation 3.4 does imply that the polarized glue is
limited in size. If the integrated polarized gluon distribution were greater
than about 0.9, the strange sea contribution in the $I^n$ column of Table II
would be positive, while the other sea flavors would be negatively polarized.
There is no apparent reason why these flavors should be polarized in a
different direction. The analysis seems to imply an inherent limit to the size
of the anomaly and thus, the size of the polarized glue.

(5) Finally, the orbital angular momentum extracted from the $J_z$ sum rule is
much smaller for all data than earlier values obtained from EMC data.\srf{18}
If the polarized gluon distribution is small enough, then both $\langle \Delta
q\rangle_{tot}$ and $\langle \Delta G\rangle$ decrease enough so that $L_z$
must be positive to account for the total spin of nucleons. Thus, both positive
and negative values for $L_z$ appear to be possible. Naturally, this opens up
the possibility that the angular momentum contribution is negligible, contrary
to the naive Skyrme model.

Clearly, these experiments have shed light on the proton spin picture. The
major unanswered questions appear to be related to the strange sea spin content
and the size of the polarized gluon distribution. These can only be reconciled
by performing other experiments which are sensitive to these quantities. To
put the strange sea picture in perspective, we can compare various results for
the polarized strange sea distributions with other models in the literature,
based on various data. These are summarized in Table III, whose
numbers represent the total contributions for each flavor (quarks and
antiquarks). The models are listed in order of increasing strange sea
contributions and refer to the experiments from which they were extracted.
The references are keyed as follows: (i) HJL- Lipkin\srf{17}; (ii) VDM- the
valence dominated model outlined in section II; (iii) GR- models presented
here (with anomaly term and higher order QCD corrections); (iv) CR- models by
Close and Roberts\srf{6}; (v) EK- recent analysis by Ellis and Karliner,\srf{7}
which incorporate the higher order QCD corrections; (vi) QRRS- models by Qiu,
{\it et. al.},\srf{8}; (vii) BEK- the model by Brodsky, {\it et. al}.\srf{18}

Disparities between these models depend on theoretical assumptions as
well as experimental data. These models can be divided into two categories:
those which satisfy the strange sea positivity constraint are listed above the
line, while those which violate this bound are below. Note that the Valence
Dominated Model and the $I^n$ data yield results that are above the line and
the proton and deuteron data do not. Thus, there is a consistency among the
proton/deuteron results (including the EMC data), which all occur above the
$Q^2$ values of the neutron results. It is possible that future data and
analysis on $I^n$ would yield a "world average" value so that it would become
more consistent with proton and deuteron data. It is clear that more tests are
necessary. As pointed out by Qiu, {\it et. al.}, and others\srf{8,27} the most
direct experiment to determine the size of the polarized sea is lepton pair
production (Drell-Yan) in polarized nucleon scattering experiments. Only then
will there be enough information to tell which assumptions about the polarized
sea are appropriate.

\tablehead{Table III: Models of the Flavor Dependence of the Polarized Sea}
$$
\begin{array}{cccc} \cr
&\underline{Model}   &\underline{\Delta u_{tot}=\Delta d_{tot}}
&\underline{\Delta s_{tot}} \cr
& HJL  (EMC)         &      -0.27                     &    +0.00 \cr
& VDM (Theory)       &      -0.09                     &    -0.01 \cr
& GR (I^n)           &      -0.10                     &    -0.02 \cr
& CR (I^n)           &      -0.12                     &    -0.03 \cr
&----------          &---------                       &----------- \cr
& GR (E143\>I^d)     &      -0.14                     &    -0.06 \cr
& GR (I^p)           &      -0.15                     &    -0.07 \cr
& GR (SMC\>I^d)      &      -0.18                     &    -0.10 \cr
& EK (SMC/E142)      &      -0.17                     &    -0.10 \cr
& CR (I^d)           &      -0.20                     &    -0.11 \cr
& CR (I^p)           &      -0.21                     &    -0.12 \cr
& QRRS (EMC)         &      -0.24                     &    -0.15 \cr
& BEK (EMC)          &      -0.26                     &    -0.23
\end{array}
$$

Thus, existing data provide valuable information regarding the proton spin
puzzle, but as a whole are not definitive in isolating the key contributions
to the proton spin. We stress that more experiments must be performed to
determine the relative contributions from various flavors of the sea and the
gluons. This is addressed in detail in the next section.

\mainhead{IV. Possible Experiments}

There are a number of experiments which are technologically feasible that
could supply some of the missing information about these distributions.
In this section, we will discuss those experiments which have been proposed
and would give specific information necessary for determining the contributions
of the sea and gluons to the overall proton spin. Detailed summaries can be
found in references 28 and 29. Table IV is extracted from reference 28 and
gives information on the spin observables which can be measured to extract the
appropriate polarized distributions. The average luminosity of these
experiments is approximately $1\cdot 10^{32}\>(cm^{-2}s^{-1})$ or greater.
Furthermore, the success of Siberian Snakes makes them all feasible. The
following discussion details the contributions and advantages which each
experiment can give in extracting the appropriate spin information.

\tablehead{Table IV: Proposed Polarization Experiments}
$$\begin{array}{cccc}
& \underline {Experiment} & \underline {Proposed\>Type}
& \underline {Measured\>Quantities} \cr
& HERMES & Deep\>Inelastic\>Scattering & A_1^p,\>g_1^p,\>\Delta q_v \cr
& SPIN   & Inelastic:\>jets & \Delta G:\>\Delta\sigma_L,\>A_{NN},\>A_{LL} \cr
& RHIC & Inelastic\>jet,\>\pi,\>\gamma & \Delta G;\>\Delta\sigma_L,\>A_{LL} \cr
& RHIC & Drell\>Yan  & \Delta S \cr
& LISS & Inelastic    & \sigma_L,\>\sigma_T,\>\Delta\sigma_L,\>\Delta G  \cr
& LHC  & Inlastic  & A_N,\>A_L,\>A_{NN},\>A_{LL} \cr
&  &  &  \cr
\end{array}
$$

\subhead{A. Extrapolation to Lower x}

The E154 and E155 deep-inelastic scattering experiments have been approved at
SLAC, with the former presently in progress. These experiments are designed to
probe slightly smaller $x$, while improving statistics and systematic errors.
The latest proposed experiment at HERA in Hamburg plans to accelerate a large
flux of polarized electrons from the storage ring and collide them with a
gaseous target.\srf{30} The HERMES detector at HERA is designed to take data
from the deep-inelastic scattering experiment at a large range of $x$ values,
down to 0.02, in contrast to 0.03 in the E143 experiment at SLAC, thus
expanding and reinforcing the SMC and E142/3 data. The gaseous target should
eliminate some of the systematic errors characteristic of solid targets,
which were used in the other experiments. With lower error bars at small $x$,
the extrapolation of the integrals should enable these experimental groups to
achieve a more accurate value for $I^p$. Thus, comparison of data to the sum
rules and the integrated polarized sea values will be more accurate.

\subhead{B. Measurement of the Polarized Sea Distributions}

The SPIN Collaboration has proposed to do fixed target $pp$ and $p\bar p$
experiments at energies of 120 GeV and 1 TeV.\srf{31} The proposal
also includes $pp$ collider experiments at 2 TeV. The luminosity at the lowest
energies would be larger than the stated average. One of the crucial
contributions that this set of experiments can make is the extremely large
range of high $p_T$ that can be covered. If this set of experiments includes
polarized lepton pair production, (Drell-Yan) then measurement of the
corresponding double spin asymmetries\srf{8,27} would give a sensitive measure
of the polarized sea distribution's size.

The Relativistic Heavy Ion Collider (RHIC) at Brookhaven is designed to
accelerate both light and heavy ions. The high energy community has
proposed that polarized $pp$ and $p\bar p$ experiments be performed, due to
the large energy and momentum transfer ranges which should be
available.\srf{32} This energy range will be covered in discrete steps of about
60, 250 and 500 GeV, but the momentum transfer range covers $0.005 \le Q^2 \le
6.0$ GeV$^2$ in a fairly continuous set of steps. There are two main proposed
detectors, STAR and PHENIX, which have different but complementing
capabilities. PHENIX is suitable for lepton detection and the wide range of
energies and momentum transfers could yield a wealth of Drell-Yan data over a
wide kinematic range. The x-dependence of the polarized sea distributions could
then be extracted to a fair degree of accuracy.

There has been considerable discussion about performing polarization
experiments at the LHC at CERN.\srf{33} Depending on the approved experiments,
there is the possibility of probing small $x$ and doing other polarized
inclusive experiments to measure both sea and gluon contributions to proton
spin. These could be made in complementary kinematic regions to those listed
in the other proton accelerators.

\subhead{C. Determination of the Polarized Gluon Distributions}

The SPIN Collaboration proposes a set of experiments, which are in the
kinematic region where the measurement of double spin asymmetries in jet
production would give a sensitive test of the polarized gluon distribution's
size.\srf{34,35} Naturally, this measurement has an effect on both $\Delta G$
and the anomaly term appearing in the polarized quark distributions.

The STAR detector at RHIC is suitable for inclusive reactions involving jet
measurements, direct photon production and pion production. All of these
would provide excellent measurements of the $Q^2$ dependence of $\Delta G$
since all are sensitive to the polarized gluon density at differing $Q^2$
values.\srf{27,35}

Recently, a proposal for a new light ion accelerator has been announced,
which will specialize in polarization experiments.\srf{36} The Light Ion
Spin Synchrotron (LISS) would be located in Indiana, and would perform a
variety of polarization experiments for both high energy and nuclear physics.
The energy range would be lower that most other experiments, complementing
the kinematic areas covered. Furthermore, both proton and deuteron beams
could be available to perform inclusive scattering experiments. They propose
to measure cross sections and longitudinal spin asymmetries which are sensitive
to the polarized gluon distribution (see Table IV).

Tests of the valence quark polarized distributions can be made, provided a
suitable polarized antiproton beam of sufficient intensity could be developed.
\srf{34} This would provide a good test of the Bjorken sum rule via measurement
of $\langle \Delta q_v\rangle$\srf{35} and the assumption of a flavor
symmetric up and down sea. This should be an experimental priority for the
spin community. Polarization experiments provide us with a unique and feasible
way of probing hadronic structure. Existing data indicate that the spin
structure of nucleons is non-trivial and has led to the formulation of a
crucial set of questions to be answered about this structure. The experiments
discussed above can and should be performed in order to shed light on the quark
and gluon spin structure of nucleons.

\newpage
\mainhead{References}
\begin{enumerate}
\item\label{R1} J. Ashman, {\it et.al.}, \pl\rf{B206}{364}{88};
\np\rf{B328}{1}{89}.
\item\label{R2} J.D. Bjorken, \pr\rf{148}{1467}{66}.
\item\label{R3} J. Ellis and R.L. Jaffe, \pr\rf{D9}{3594}{74}.
\item\label{R4} B. Adeva, {\it et. al.}, \pl\rf{B320}{400}{94}; D. Adams,
{\it et. al.}, \pl\rf{B329}{399}{94} and \pl\rf{B357}{248}{95}.
\item\label{R5} P.L. Anthony, {\it et. al.}, \prl\rf{71}{959}{93}; K. Abe,
{\it et. al.}, \prl\rf{74}{346}{95}.
\item\label{R6} F.E. Close and R.G. Roberts, \pl\rf{B316}{165}{93}.
\item\label{R7} J. Ellis and M. Karliner, \pl\rf{B341}{397}{95}.
\item\label{R8} J.-W. Qiu, G.P. Ramsey, D.G. Richards and D. Sivers,
\pr\rf{D41}{65}{90}.
\item\label{R9} R. Carlitz and J. Kaur \prl\rf{38}{673}{77}.
\item\label{R10} M. Gluck, E. Reya and A. Vogt, \zp\rf{C48}{471}{90};
\zp\rf{C53}{127}{92}; \pl\rf{B306}{391}{93} and A. Vogt, \pl\rf{B354}{145}{95}.
\item\label{R11} A.D. Martin, R.G. Roberts and W.J. Stirling, J. Phys. G
\rf{19}{1429}{93}; \pl\rf{B306}{145}{93}; \pr\rf{D50}{6734}{94}; \pl\rf{B354}
{155}{95}.
\item\label{R12} S.J. Dong, J.-F. Laga$\ddot{e}$ and K.F. Liu, preprint UK-01
(1995).
\item\label{R13} B.-A. Li, {\it et. al.}, \pr\rf{D43}{1515}{91}.
\item\label{R14} S.A. Larin, F.V. Tkachev and J.A.M. Vermaseren, \prl\rf
{66}{862}{91} and S.A. Larin and J.A.M. Vermaseren, \pl\rf{B259}{345}{91};
A.L. Kataev and V. Starshenko, CERN-TH-7198-94.
\item\label{R15} E. Stein, {\it et. al.}, \pl\rf{B353}{107}{95}.
\item\label{R16} R.L. Jaffe \pl\rf{B193}{101}{87}.
\item\label{R17} H.J. Lipkin, \pl\rf{B214}{429}{88} and Weizmann Preprint
WIS-94/24/May-PH (1994).
\item\label{R18} F.E. Close and R.G. Roberts, \prl\rf{60}{1471}{88}; S.J.
Brodsky, J. Ellis and M. Karliner, \pl\rf{B206}{309}{88}; M. Anselmino, B.L.
Ioffe, and E. Leader Yad. Fiz. \rf{49}{214}{89}; H.J.Lipkin, \pl\rf{B256}{284}
{91}.
\item\label{R19} G. Preparata, P.G. Ratcliffe and J. Soffer, \pl\rf{B273}{306}
{91}
\item\label{R20} G. Ramsey, J.-W. Qiu, D.G. Richards and D. Sivers,
\pr\rf{D39}{361}{89}.
\item\label{R21} A.V. Efremov and O.V. Teryaev, JINR Report E2-88-287 (1988);
G. Alterelli and G.G. Ross, \pl\rf{B212}{391}{88}; R.D. Carlitz, J.C. Collins,
and A.H. Mueller, \pl\rf{B214}{229}{88}.
\item\label{R22} S.J. Brodsky, M. Burkardt and I. Schmidt, \np\rf{B441}{197}
{95}; P. Chiapetta and J. Soffer, \pr\rf{D31}{1019}{85}; M. Einhorn and
J. Soffer, \np\rf{B274}{714}{86}
\item\label{R23} M. Derrick, {\it et. al.}, \zp\rf{C65}{379}{95} and
T. Ahmed, {\it et. al.}, \pl\rf{B348}{681}{95}.
\item\label{R24} S. Platchkov, preprint DAPNIA SPhN 93 53: invited talk at the
14th Eur. Conf. on Few Body Problems, Amsterdam, 1993; M Lacombe, {\it et.
al.}, \pl\rf{B101}{139}{81}.
\item\label{R25} S.A. Larin, \pl\rf{B334}{192}{94}.
\item\label{R26} B. Ehrnsperger and A. Sch$\ddot{a}$fer, \pl\rf{B348}{619}{95}.
\item\label{R27} E.L. Berger and J.-W. Qiu, \pr\rf{D40}{778}{89}; P.M.
Nadolsky, Z. Phys. \rf{C62}{109}{94}; H.-Y. Cheng and S.-N. Lai,
\pr\rf{D41}{91}{90} and D. deFlorian, {\it et. al.}, \pr\rf{D51}{37}{95}.
\item\label{R28} G.P. Ramsey, Particle World, \rf{4}{No. 3}{95}.
\item\label{R29} S.B. Nurushev, IHEP preprint IHEP 91-103, Protvino, Russia.
\item\label{R30} See, for example, Physics Today, \rf{47 (No. 11)}{19}{94}.
\item\label{R31} SPIN Collaboration, A.D. Krisch, {\it et. al.}, "Expression
of Interest: Accelerated Polarized Beam Experiments at the Fermilab Tevatron",
May, 1994.
\item\label{R32} RHIC Spin Collaboration, "Proposal on Spin Physics Using the
RHIC Polarized Collider", August, 1992; and STAR-RSC Update, October, 1993.
\item\label{R33} LHC Polarization Proposal, 1994.
\item\label{R34} G.P. Ramsey, D. Richards and D. Sivers \pr\rf{D37}{314}{88}.
\item\label{R35} G.P. Ramsey and D. Sivers, \pr\rf{D43}{2861}{91}.
\item\label{R36} J.M. Cameron and S. Vigdor, "The LISS Brief", proceedings
of the XI International Symposium on High Energy Spin Physics, Bloomington, IN,
September, 1994; c 1995, AIP.
\end{enumerate}
\end{document}